# Photoinduced coherent acoustic phonon dynamics inside Mott insulator $Sr_2IrO_4$ films observed by femtosecond X-ray pulses

Bing-Bing Zhang,[1] Jian Liu,[2] Xu Wei,[1] Da-Rui Sun,[1] Quan-Jie Jia,[1] Yuelin Li[3,a)] and Ye Tao[1,b)]

[1]*Institute of High Energy Physics, Chinese Academy of Science, Beijing, 100049, China*

[2]*Department of Physics and Astronomy, University of Tennessee, TN 37996, Knoxville, USA*

[3]*X-ray Science Division, Argonne National Laboratory, Lemont, IL 60565, USA*

---

a) Electronic mail: ylli@aps.anl.gov
b) Electronic mail: taoy@ihep.ac.cn

**ABSTRACT**

We investigate the transient photoexcited lattice dynamics in a layered perovskite Mott insulator $Sr_2IrO_4$ by femtosecond X-ray diffraction using a laser plasma-based X-ray source. Ultrafast structural dynamics of $Sr_2IrO_4$ thin films are determined by observing the shift and broadening of the (0012) Bragg diffraction after excitation by 1.5 eV and 3.0 eV pump photons for films with different thicknesses. The observed transient lattice response can be well interpreted as a distinct three-step dynamics due to the propagation of coherent acoustic phonons generated by the photoinduced quasiparticles (QP). Employing a normalized phonon propagation model, we found that the photoinduced angular shifts of the Bragg peak collapse into a universal curve after introducing a normalized coordinates to account for different thicknesses and pump photon energies, pinpointing the origin of the lattice distortion and its early evolution. In addition, a transient photocurrent measurement indicates that the photoinduced QPs are charge neutral excitons. Mapping the phonon propagation and correlating its dynamics with the QP by ultrafast X-ray diffraction (UXRD) establish a powerful way to study electron-phonon coupling and uncover the exotic physics in strongly correlated systems under nonequilibrium conditions.

**Manuscript text**

The layer perovskite Sr2IrO4 was found to host antiferromagnetic Mott insulating state due to its comparable energy scale among electron bandwidth, Coulomb repulsion and spin-orbital coupling energy.[1] In addition, the spin-orbital coupled material exhibits plenty of structural and electronic similarities to the superconducting parent phase of the cuprates $La_2CuO_4$, and high-temperature superconductivity have been theoretically predicted in doped $Sr_2IrO_4$.[1-3]

Electron-phonon coupling plays an essential role in understanding the pairing mechanism of unconventional superconductivity systems.[4-6] Photo-exciting these systems by laser provides a powerful method to interrogate and even manipulate the exotic properties in non-equilibrium states.[7-10] Photoinduced structural dynamics and its direct involvement with electron-phonon coupling have been determined in $La_2CuO_4$ using time-resolved electron diffraction and pump-probe absorption spectroscopy.[11,12] Strong electron-phonon coupling and temperature dependence of electronic structure of $Sr_2IrO_4$ have been studied by laser pump-laser probe spectroscopy experiments.[13,14] Recently, magnetic resonant inelastic X-ray scattering has been performed at free-electron laser to determine the transient magnetic dynamics in photo-doped $Sr_2IrO_4$.[15] By combining synchrotron based UXRD and transient optical absorption spectroscopy, strong photoinduced lattice correlation has been found in photo-doped $Sr_2IrO_4$ and attributed to the presence of photoinduced QPs.[16] However, the temporal resolution limit of SR precludes tracking the dynamics within the first 100 picosecond (ps) after excitation, which is critical for understanding the QP-phonon coupling mechanism.[17]

In this letter, we trace the dynamics of the lattice response via UXRD at a laser plasma-based X-ray source[18] by probing the transient shift and broadening of $Sr_2IrO_4$ film (0012) Bragg peak after excitation with sub-ps time resolution. Results are obtained for films with different thicknesses and excitation photon energies. The photoinduced strain evolution is found to incorporate a fast rise timescale (tens of ps) and two consecutive recovery timescales. We attribute the transient dynamics within the first few tens of ps to the propagation of coherent acoustic phonons arising from the sudden generation of photoinduced QPs in the films. An analytical phonon propagation model, originally developed to describe the propagation of strain arising from thermally induced coherent phonons, is adapted to quantify the ultrafast evolution.[16,19] Our transient photo-current measurements also unambiguously reveal the nature of the QPs as charge neutral entities. This work provides insight into the atomic scale dynamic process in $Sr_2IrO_4$ with sub-ps temporal resolution.

The samples are 50 and 100nm thick $Sr_2IrO_4$ films grown on (001) $SrTiO_3$ substrate using pulsed laser deposition.[20] The pulse duration of the generated Cu-Kα X-rays is estimated to be 200~300 femtosecond (fs). 1 kHz pump laser pulses, split from the same laser system for the X-ray generation, are impinged on the sample with photon energies of 1.5 eV and 3.0 eV. Using Montel X-ray optics, the X-ray beam is focused onto the sample with a diameter of approximately 300 μm (FWHM), overfilled by the 1 mm (FWHM) pump spot at a fluence of 8mJ/cm2. The time delay between the pump pulse and probe X-ray pulse is adjustable via an optical delay translation stage. The scattered X-ray signals are gathered by a 2D detector (Pilatus 100K).

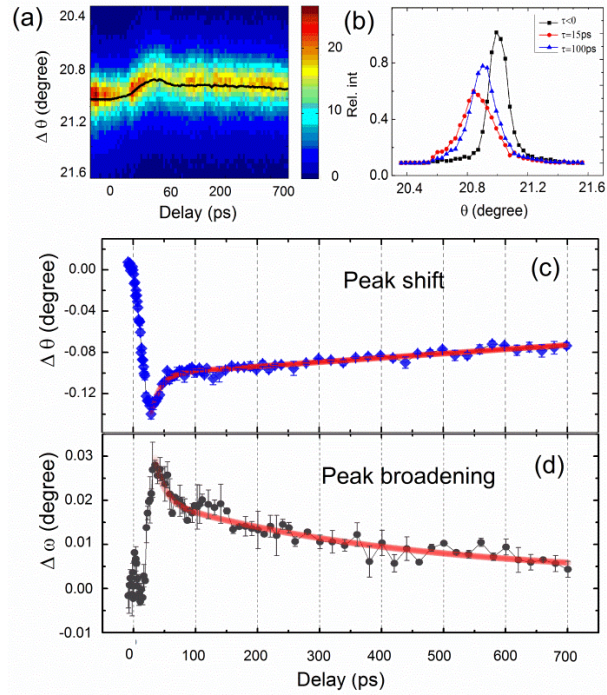

FIG. 1. Time resolved rocking intensity of the (0 0 12) Bragg reflection from a 100 nm $Sr_2IrO_4$ sample, with 1.5 eV pump photon energy. (a) The rocking curve as a function of time (delay between X-ray and the laser pulse), where θ is the scattering angle. (b) Rocking curves at 15 ps and 100 ps and (c) Shift and (d) broadening of the rocking curve as a function time. The red solid lines in (c) and (d) are the biexponential fit.

Fig. 1 shows the evolution of the (0012) Bragg peak of the 100 nm $Sr_2IrO_4$ sample at 1.5 eV photoexcitation at room temperature. Fig. 1(a) illustrates the transient rocking curves from -10~700 ps. Upon photoexcitation, the reflection shifts to a lower Bragg angle, as shown in Fig. 1(b), indicating the average out-of-plane lattice expansion. The subsequent relaxation dynamics can be well fitted by a biexponential decay function with a fast time constant of 15 ps and a slow time constant of 15 ns, as shown in Fig. 1(c). Meanwhile, the broadening of the Bragg peak illustrated in Fig.1 (d) exhibits a similar decay behavior, suggesting the inhomogeneous atomic movements after excitation. Unlike the results from synchrotron source, 16 current sub-ps XRD reveal the hidden details of the onset and initial evolution of photoinduced strain. Note that a linear dependence of the maximum strain on the laser fluence was measured up to 20 mJ/cm2 in previously.

The different absorption coefficients, thus the optical penetration depth ζ, at these different excitation photon energies (1.5 eV~70 nm, 3.0 eV~30 nm) lead to very different dynamics shown in Fig. 2. We note that, although the recoveries of the photoinduced angular shifts suggest almost the same biexponential decay for all cases, as represented by the results of the 100 nm sample (Fig. 2(a)), their peak broadening evolutions appear to be more complicated. As shown in Fig. 2(c), the 1.5 eV-50 nm case presents a smaller photon-induced broadening followed by a relatively gentle decay, while the other cases show rapid narrowing after reaching the maximum broadening, displaying distinct biexponential decay.

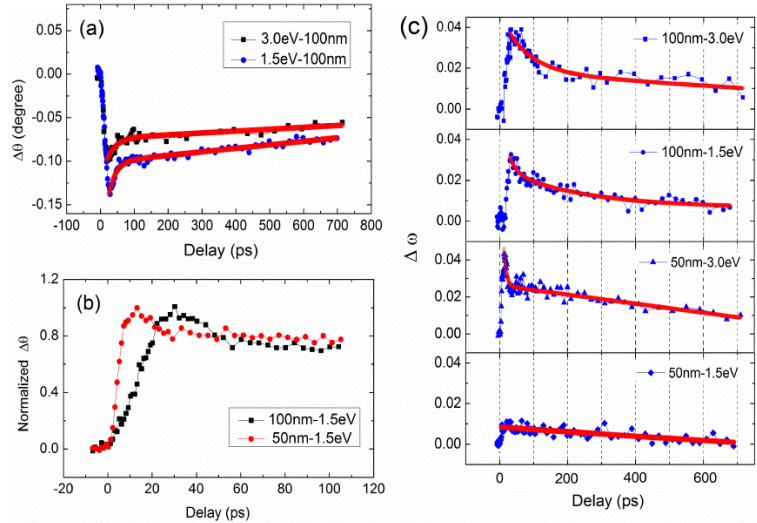

FIG. 2. (a) Diffraction peak shifts of 100 nm sample upon 1.5 and 3.0 eV pump photon energy, the red solid lines are biexponential fits. (b) The evolution of normalized shift of the rocking curve in the first 100 ps range for two different films after 1.5 eV laser pump. (c) The broadenings of the reflection on 50 nm and 100 nm samples at different pump photon energy (symbols), the red solid lines are biexponential fits except for the 50nm-1.5eV case, where it is to guide the eye.

Similar to previous results,[16] thicker sample possesses longer recovery time both for the peak shift and broadening as shown in Fig. 2(b) and 2(c). More importantly, we find the quantitative temporal relation: it takes the 100 nm film (~27.5±1 ps) twice as long as the 50nm sample （~13.7±0.5 ps） to reach the maximum angular shift (Fig. 2(b)). The same is applicable to the fast relaxation from the maximum shift to the sub-equilibrium state. In fact, the introduction of a new time coordinate $\tau=\upsilon*t/d$ (where $t$ is delay time, $d$ is film thickness, and $\upsilon$ denotes the sound velocity normal to the surface, respectively) eliminates the effect of the film thickness and all the strain dynamics collapse into one curve (Fig. 3(a)), similar to the recent work on thermal stress effects in other thin films.[19]

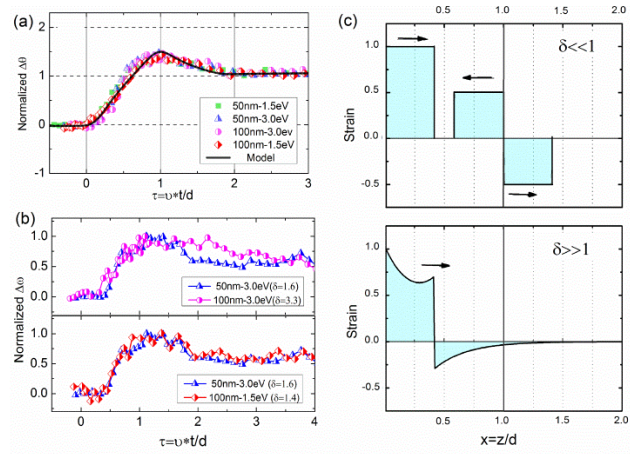

FIG. 3. (a) The normalized peak shift under different conditions, using a normalized coordinate, the black solid line shows the normalized average strain (δ=1.6) basing on the analytical model, small deviations of normalized strain among different δ values could be found in Fig. 1S. (b) Normalized peak broadening for three δ>1 cases , the 50nm-3.0eV and 100nm-1.5eV cases match with each other due to their similar δ value. (c) The normalized strain for two limiting case in a normalized x coordinate at τ=0.4.

As reported in the previous study,[16] the origin of the photoinduced strain is attributed to the lattice distortion correlated to the presence of QPs, where the magnitude of the strain is related to the local QP density. The shift and the broadening of the diffraction peak are directly proportional to the average and the spatial variation of the strain in the thin film[16]: $-\Delta q_z(t) \propto \langle \eta(z,t) \rangle_z$ and $\Delta w(t) \propto \Delta_z \eta(z,t)$ [17,19]. Here we interpret the onset of the overall lattice dynamics as due to the propagation of coherent acoustic phonons in the thin film initialized by the generation of QPs. We introduce a factor $\delta=d/\zeta$ [19] to describe the initial excitation spatial profile, where $\zeta$ denotes the penetration depth of the pump laser. For δ<<1, stress is homogeneous and the initial tensile strain waves are endowed with different travelling directions.[19] However, for δ>>1, the laser only penetrates a very thin surface layer of the sample, generating inhomogeneous stress and bipolar strain in the film (Fig. 3(c)).

Different δ leads to different strain profile η(x, τ), thus different dynamics in angular shift and broadening of Bragg peaks. This has made it possible to untangle the physics that governs the diffusion dynamics of the quasiparticles. Generally, for δ>1, the Bragg peak splits as a consequence of the coexistence of the tensile and compressive strained regions within the film. However, due to the large peak width resulting from the small film thickness, only Bragg peak broadening could be observed. The larger the δ is, the more inhomogeneous the strain distribution is, thus a bigger broadening. Then the 3.0 eV case shows larger peak width change than that in the 1.5 eV case, as shown in Fig. 2(c), identical to the previous results.[16] Moreover, as shown in Fig. 3(b), upon introducing the normalized coordinate τ, the peak broadening evolution of 50 nm-3.0 eV (δ=1.6) almost coincides with that of the 100 nm-1.5 eV (δ=1.4) due to their comparable δ value, but not the 100 nm-3.0 eV case (δ=3.3).

Based on the analytical phonon propagation model,[19] the calculated average strain ⟨η(x, τ)⟩ reproduces the measured temporal evolution, as shown in Fig. 3(a). A characteristic ratio of 1.5 between the strain at τ=1, when the strain front travels once in the film, and τ=2, when the strain travels back and forth in the film, is also well reproduced. In Fig. 3(a), a convoluted 250 fs Gaussian X-ray pulse for the UXRD setup is used. The fitting value of the sound velocity in iridates is $\upsilon=d/t \sim 3.65(\pm 0.15)$ nm/ps, in good agreement with calculated values of 3.99 nm/ps (S1.2).

Our sub-ps UXRD experiments overcome the limitation of SR pulse duration and probe the emergence and evolution of the transient strain within the first 100 ps. It is clear that upon photoexcitation, an instantaneous strain is generated at the film surface and interface followed by propagation of acoustic phonons leading to the early development of the peak shift for τ<2. Further relaxation processes (τ>2) can be modeled by the diffusion of carriers and their recombination or dissociation at the film surface and interface.[16]

Finally, to understand the nature of the quasiparticles, a transient photocurrent measurement is performed with 1.5 eV photoexcitation. Two electrodes 0.5 mm apart were made on a 100 nm $Sr_2IrO_4$ film. A fast current amplifier (Femto DHPCA-100) was employed to apply a bias voltage ranging from 0.2 to 9 V to the sample. The transient photo currents were measured at different bias voltages, using a fast 16 GHz oscilloscope (DSAX91604A Infiniium). At constant laser irradiance, we found that the peak amplitude of the photocurrent is proportional to the bias voltage (the inset of Fig. 4). Importantly, the dynamics of the photocurrent is independent of the bias voltage (Fig. 4) with a rise time of about 20 ns (limited by the current amplifier's temporal resolution) and a decay time of about 100 ns.

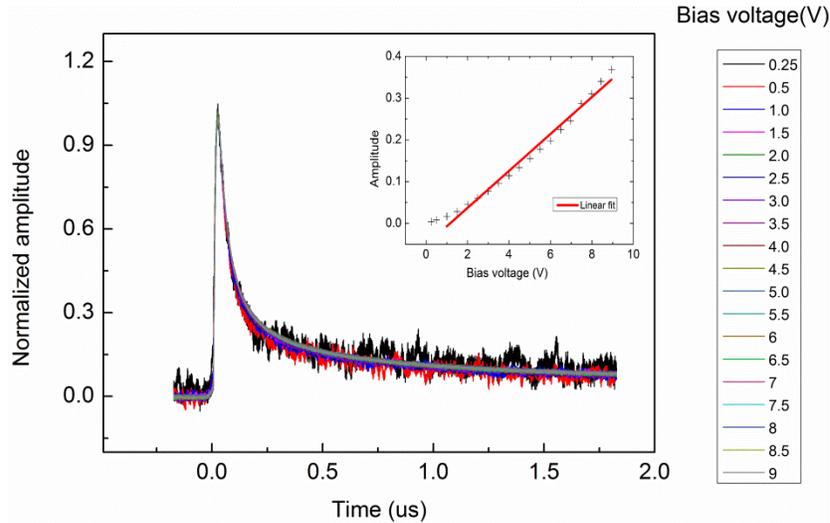

FIG. 4. Normalized photocurrent dynamics of $Sr_2IrO_4$ at different bias voltages (0.25, 0.5, 1.0 … 8, 8.5 and 9V), the inset shows peak amplitude of photocurrent as a function of bias voltages.

The bias-independent dynamics cannot be caused by local trapping of the QPs as this would give a QP dynamics independent of the film thickness, contradicting the result in Fig. 2 and in Ref.[16] Instead, the effect points to the charge neutrality nature of the photoinduced QPs; otherwise, a higher bias would lead to a faster decay time due to the higher mobility. The charge neutrality of the QPs is a strong indication that they have a bound electron-hole configuration, likely the Hubbard exciton,[21] with an electron at the bottom of the upper Hubbard band (doublon) and a hole at the top of the lower Hubbard band (holon). In this case, the higher amplitude at higher bias is due to the more efficient charge extraction at the electrode/film interfaces, where the electrons and holes separate due to local band bending.

We have studied the photoexcitation dynamics in $Sr_2IrO_4$ film by means of sub-ps XRD and nanosecond photocurrent measurements. We traced the emergence and evolution of photoinduced atomic displacement by UXRD. The broadening of the corresponding Bragg peak indicates the inhomogeneous spatial strain profile. The photocurrent dynamics measurements suggest that photoinduced QPs are likely charge neutral Hubbard excitons. The observation of coherent acoustic phonons demonstrates the large electron-phonon coupling that persists in iridates after photoexcitation, offering a unique route to understand the complex coupling between different degrees of freedoms in strongly correlated systems, including HTSC systems.

## SUPPLEMENTARY MATERIAL

See supplementary material for detailed description of the phonon propagation model and the calculation of sound velocity.

## ACKNOWLEDGMENTS

This work is supported by the National Science Foundation of China (U1332205). J. L. is sponsored by the Science Alliance Joint Directed Research & Development Program and the Transdisciplinary Academy Program at the University of Tennessee. Y. L is sponsored by the Advanced Photon Source, a U.S. Department of Energy (DOE) Office of Science User Facility operated for the DOE Office of Science by Argonne National Laboratory under Contract No. DE-AC02-06CH11357. The photocurrent measurement was performed using the APS-7ID kHz pump laser system.